\documentclass[conference]{IEEEtran}
\IEEEoverridecommandlockouts
\usepackage{cite}
\usepackage{amsmath,amssymb,amsfonts}
\usepackage{algorithmic}
\usepackage{graphicx}
\usepackage{textcomp}
\usepackage{xcolor}
\usepackage{threeparttable}
\usepackage{tabularray}
\usepackage{xurl}
\def\BibTeX{{\rm B\kern-.05em{\sc i\kern-.025em b}\kern-.08em
    T\kern-.1667em\lower.7ex\hbox{E}\kern-.125emX}}
\begin{document}

\title{Sustainable broadcasting in Blockchain Networks with Reinforcement Learning}

\author{\IEEEauthorblockN{Danila Valko}
\IEEEauthorblockA{\textit{L3S Research Center, Hannover, Germany} \\
danila.valko@l3s.de}
\and
\IEEEauthorblockN{Daniel Kudenko}
\IEEEauthorblockA{\textit{L3S Research Center, Hannover, Germany} \\
kudenko@l3s.de}
}

\maketitle


\begin{abstract}
Recent estimates put the carbon footprint of Bitcoin and Ethereum at an average of 64 and 26 million tonnes of $CO_{2}$ per year, respectively. To address this growing problem, several possible approaches have been proposed in the literature: creating alternative blockchain consensus mechanisms, applying redundancy reduction techniques, utilizing renewable energy sources, and employing energy-efficient devices, etc. In this paper, we follow the second avenue and propose an efficient approach based on reinforcement learning that improves the block broadcasting scheme in blockchain networks. Such an improvement concerns both the block propagation time and the number of messages to be broadcast before the network reaches consistency. The latter determines the amount of traffic and distributed energy consumption across the blockchain network infrastructure.

In particular, we implemented a reinforcement learning (RL) agent to efficiently re-prioritize the broadcast order in the default blockchain block propagation protocol based on real-time transport layer network information. This approach reduces the average propagation time as well as the number of messages needed to reach network consistency. Since blockchain networks are highly distributed around the world, this seemingly small improvement could make a big difference in the context of overall energy consumption and the network's impact on the environment.

To train the agent and validate the approach, we used a blockchain simulator that was extended by adding a broadcast monitoring interface and was integrated into the RL environment. We then ran a series of simulations and general statistical analyses to compare the performance of the default block propagation scheme and the scheme with the RL agent involved. The analysis and experimental results confirmed that the proposed improvement of the block propagation scheme could cleverly handle network dynamics and achieve better results than the default approach. Additionally, our technical integration of the simulator and developed RL environment can be used as a complete solution for further study of new schemes and protocols that use RL or other ML techniques.
\end{abstract}

\maketitle

\section{Introduction}

\begin{figure}
    \centering
    \includegraphics[width=1.0\linewidth]{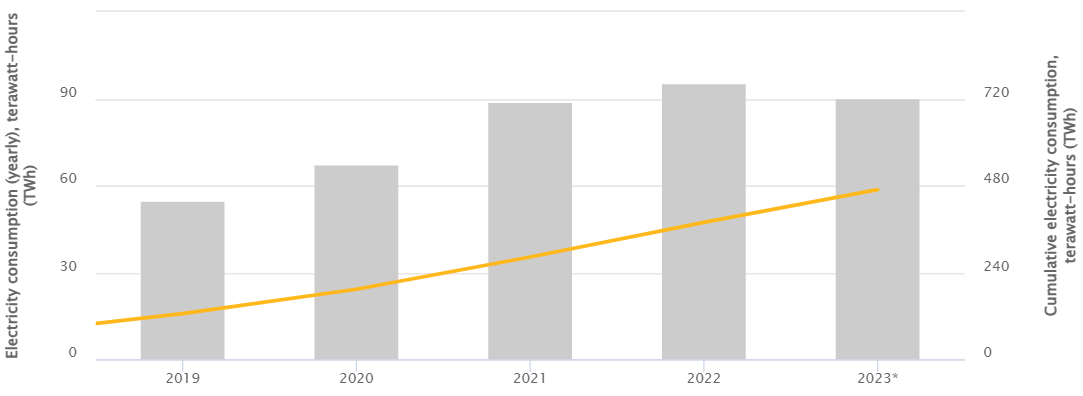} (a)
    \includegraphics[width=1.0\linewidth]{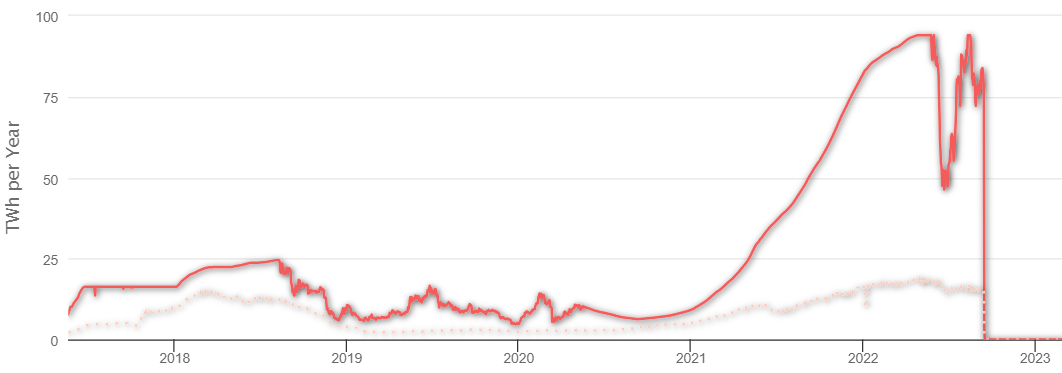} (b)
    \caption{The Energy Consumption Index of Bitcoin and Ethereum}
    \label{consumption}
    \begin{tablenotes}
      \small
      \item Note. It shows yearly energy consumption of Bitcoin (a) and Ethereum (b) infrastructure measured in terawatt-hours (TWh), data was obtained from https://ccaf.io/cbnsi/cbeci and https://digiconomist.net/ethereum-energy-consumption.
    \end{tablenotes}
\end{figure}

Bitcoin and Ethereum are the most carbon-intensive blockchain-based P2P payment networks. They consume a growing amount of energy (Fig. \ref{consumption}) and emit an average of 64 and 26 million tons of $CO_{2}$ per year, respectively\cite{Kohli2023, Guidelines2023}. Besides Bitcoin and Ethereum, several other large blockchains produce on average more than 750 tons of $CO_{2}$ equivalent each year\cite{Guidelines2023}. Consequently, new environmental, social, and economic challenges confront inventors, researchers, and governments\cite{Guidelines2023}. Thus, every small improvement in blockchain protocols, schemes, and algorithms may lead to significant spatiotemporally distributed effects. Despite this, research in this direction still does not explicitly discuss the environmental consequences of proposed inventions and algorithmic improvements.

To address this growing problem, several possible approaches for blockchain networks have been proposed in the literature: creating alternative consensus mechanisms, applying redundancy reduction techniques, and utilizing renewable energy sources and energy-efficient devices\cite{Kohli2023}. In this paper, we follow the second avenue and propose an original and efficient approach based on reinforcement learning (RL) that improves the block broadcasting scheme in blockchain-based networks.

The use of RL-based approaches to solve routing problems in different types of networks has been widely discussed in the literature (see review \cite{Mammeri2019}), but they do not directly address global peer-to-peer networks such as Bitcoin and Ethereum (see details in \textsection\ref{relwork}). In addition, we want to emphasize the environmental impact of the proposed solution and therefore choose and discuss relevant metrics as a proxy for the environmental impact.

The limitations of existing blockchain routing schemes were recently addressed in \cite{Yang2019}. They showed that RL could enhance routing adaptability by dynamically selecting more dependable and efficient routes, but in the context of wireless sensor networks. We extended this concept to Ethereum-related networks, proving that a RL agent can effectively improve broadcasting order.

As we intended to perform cost-effective, controlled, and reproducible experimentation, our approach was tested on the blockchain simulator developed by Faria et al.\cite{Faria2019}, which was extended by implementing the default Ethereum broadcasting mechanism. We validated the proposed scheme improvement through a number of simulations and demonstrated that the approach reduces the average block propagation time while maintaining a good blockchain synchronization rate. By implementing proposed improvements the total carbon emissions could be reduced by approximately 0.331 gCO2eq. during each broadcasting step. Additionally, the reduction in synchronization time should also lead to a lower energy consumption and carbon emissions.

\section{Background and related work}
\subsection{Blockchain network and block propagation}
It is well known that the blockchain concept was introduced with the Bitcoin cryptocurrency system by Satoshi Nakamoto in 2008 \cite{nakamoto2009}. The blockchain can be regarded as an immutable and decentralized database maintaining a continuously growing list of ordered records, called blocks, that are secured from tampering and shared among participating members. The blockchain is an important approach for open environments \cite{Alam2022}, as it is dedicated to memorizing data, executing transactions, performing functions, and providing trust and secure computations \cite{Zhang2019}.

Bitcoin and Ethereum are the most common blockchain cryptocurrency architectures built on an unstructured peer-to-peer (P2P) network model. The P2P architecture of blockchain allows all cryptocurrencies to be transferred worldwide without the need for any middleman, intermediaries, or central server. With the distributed P2P network, anyone who wishes to participate in the process of verifying and validating blocks can set up a node \cite{Sharma2022}.

New blocks in Bitcoin and Ethereum can only join the chain when other nodes validate these blocks, which occurs after executing a decentralized consensus procedure \cite{Lin2020}. The number of blocks increases over time, and these blocks are linked together using cryptography to form a chain. Each individual block holds a cryptographic hash of the antecedent one.

When a node initializes, it attempts to discover a set of peers to establish outgoing or incoming transport layer network connections that are based on the TCP protocol. These TCP connections are used for transaction and block propagation. Each node maintains a list of peers' IP addresses. For instance, according to the default protocol in the Bitcoin Core project \cite{BProject2020}, an average node in the network initiates up to 8 outgoing connections and accepts up to 117 incoming connections. Supernodes may establish more than 8 outgoing connections \cite{Zhang2021}.

\begin{figure}
    \centering
    \includegraphics[width=1.0\linewidth]{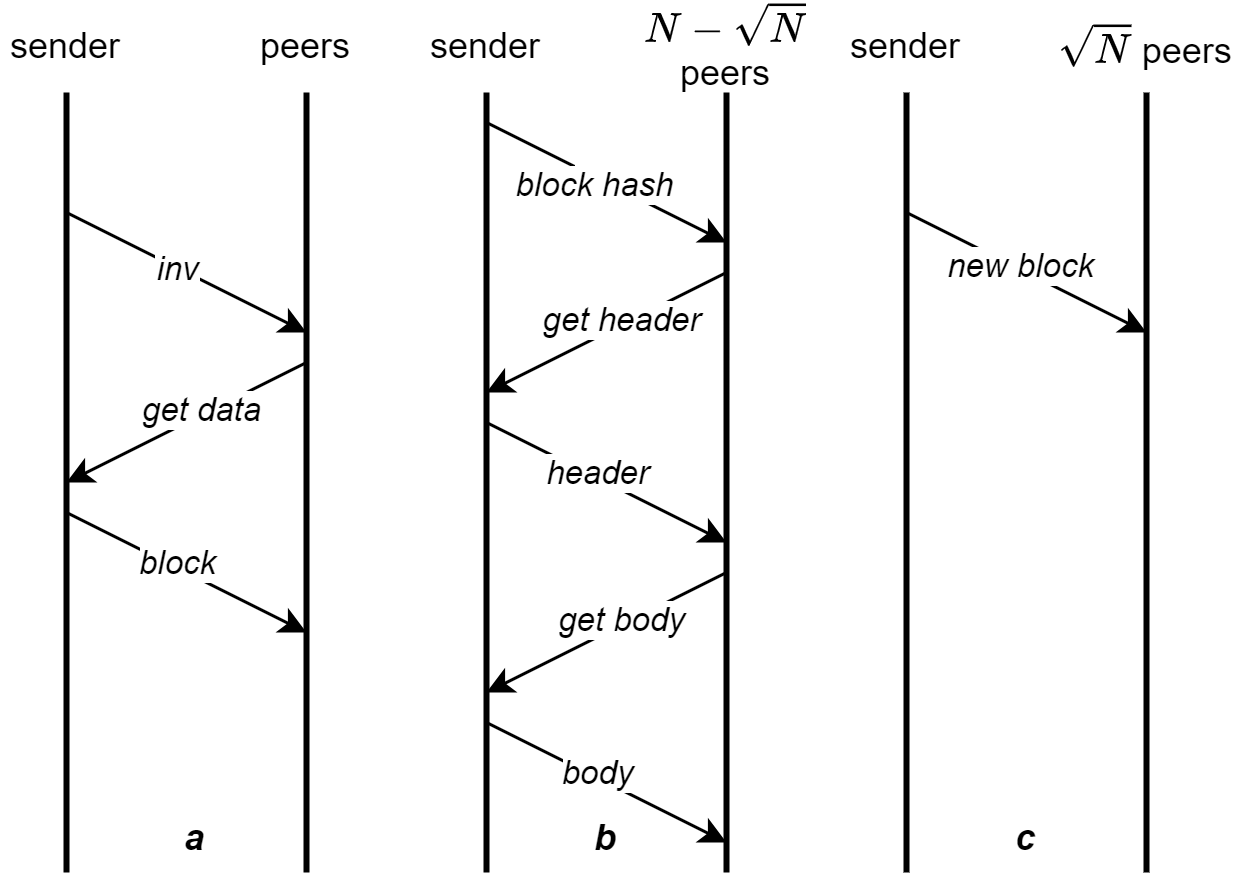}
    \caption{Standard block-hash propagation scheme (BPP)~\cite{Ma2022}}
    \label{fig1}
    \begin{tablenotes}
      \small
      \item Note. a – default propagation scheme in Bitcoin; b, c – default propagation scheme in Ethereum.
    \end{tablenotes}
\end{figure}

The standard block-hash propagation (BPP) scheme (protocol) in the Bitcoin/Ethereum blockchain implies that a sending node forwards a new block to its $N$ neighbor nodes \cite{Zhao2022}. In Bitcoin, when a certain block or transaction arrives and is verified by a node, it notifies the neighbors using an inventory ($inv$) message to let them know that a new block or transaction is available and ready to send. The $inv$ message consists of the hash of the mentioned block or transaction. When a node receives such a message for a block or transaction that it has not already seen, it replies to the $inv$ message with a "get data" message. Upon receiving the "get data" message, the node transfers the block or transaction to the sender of this message (Fig. \ref{fig1}, a). In Ethereum, the sending node randomly selects $\sqrt{N}$ neighbor nodes to forward the full block directly after verifying the block head information. It then announces the block hash to the remaining neighbor nodes after verifying the full block. The neighbor nodes that do not have the block will request the block header and block body successively from this sending node to reconstruct the full block (Fig. \ref{fig1}, b, c). 

Message broadcasting and traffic load are the most important components that determine energy consumption in most networks, including blockchain-related networks. However, a significant step towards sustainability in blockchain-related networks was taken with the invention of the proof-of-stake (PoS) consensus algorithm \cite{King2012}. The idea for PoS originated as a way to create a less energy-consuming alternative to Bitcoin's proof-of-work algorithm (PoW), which requires miners to solve cryptographic puzzles to verify transactions on the blockchain and involves substantial computations leading to high energy consumption. Nevertheless, due to the expansion of the blockchain infrastructure and the increasing number of transactions, further improvements are greatly needed \cite{Kohli2023}.

\subsection{Reinforcement learning for distributed networks}

Reinforcement learning has shown excellent ability in solving some complex problems, including those related to blockchain \cite{Bai2022}. It uses rewards to enable the algorithm to continuously optimize decision-making in the learning process, thereby learning an optimal mapping from state to action \cite{Nian2020}. The rewards for performing actions are delayed; that is to say, the pros and cons of the current action cannot be judged immediately. Only after the action has an impact on the state, the cumulative reward obtained from the execution of the action up to a certain moment afterward is calculated. Then, the model accomplishes the optimization of the action by reward \cite{Hatem2009}.

Basically, a task in RL is modeled as a Markov Decision Process (MDP). An MDP, denoted as $M = (S, A, T, R, \gamma)$, where $S$ and $A$ are the state and action space respectively. $T(s, a, s') : S \times A \times S \rightarrow [0, 1]$ is the probability of reaching state $s'$ from state $s$ after executing action $a$. The reward function $R(s, a, s') : S \times A \times S \rightarrow \mathbb{R}$ assigns a numerical reward to a state transition from $s$ to $s'$ with respect to the executed action $a$. A policy $\pi(s, a) : S \times A \rightarrow [0, 1]$ defines how the agents should act in the environment through the probability distribution over all actions in every state. The decay factor $\gamma \leq 1$ is used to define the expected discounted return $D_t = \sum_{k=0}^{\infty} \gamma^k R_{t+k}$ and the value function $V(s) = E_{A_{t} \sim \pi (S_t)} [\sum_{t=0}^{\infty} \gamma^t R_t | S_0 = s]$. The RL agent learns a policy that maximizes the expected discounted return, where the optimal policy has the maximum expected discounted return~\cite{Sutton1998}.

The use of the RL-based approach to solve routing problems in different types of networks has been widely discussed in the literature (see for review \cite{Mammeri2019}). For example, Boyan et al. \cite{Boyan1993} first combined the Q-learning algorithm with packet routing to dynamically learn the routing situation and find the shortest path in the network. Gao et al. \cite{Gao2017} proposed a multi-agent routing algorithm with Q-learning and backpressure, where each routing node needs only local information about the neighbor routing nodes to solve this problem. Mayadunna et al. \cite{Mayadunna2017} and Yang et al. \cite{Yang2019} proposed RL-based malicious routing node detection schemes for mobile ad-hoc networks and wireless sensor networks, respectively.

\subsection{Related work and sustainability concern}
\label{relwork}
Several schemes and protocols have been proposed to reduce propagation time and related traffic load in blockchain-based networks, while also ensuring network security: compact block propagation \cite{Zhang2021}, hybrid compact block propagation \cite{Zhao2022}, bodiless block propagation \cite{Zhao2022bodyless}, etc. These proposals mainly focus on algorithms and technical improvements of the Bitcoin or Ethereum core protocol itself. Such enhancements often necessitate significant changes in network design and infrastructure, and in some cases, even a fork of Bitcoin/Ethereum. Therefore, our work emphasizes the network transport layer, which is typically invariant of the underlying top-level protocols and requires relatively small changes in client software. By slightly modifying the previously mentioned default propagation procedure (BPP) at the broadcasting step with a RL-based approach, technically, such an approach can be implemented with any other propagation scheme that usually does not depend on the transport layer of the network.

As mentioned above, recent works consider RL implementations for blockchain-based networks to improve their methods and algorithms. For instance, in \cite{Alam2022}, by combining RL and blockchain for IoT networks, a decentralized communication structure for scalable and trustworthy information allocation was developed (for a review on RL for IoT, see \cite{Gasmi2023, Gao2021}). In \cite{Gadiraju2023}, deep RL was used to optimize the performance of the recently invented Prism proof-of-work blockchain protocol \cite{Bagaria2019} and enhance the number of votes without violating security and latency performance guarantees \cite{Gadiraju2023}. In \cite{Li2023}, RL was used to auto-tune network fabric in permissioned blockchain systems and demonstrated an ability to identify optimal network configurations.

There are also a number of papers devoted to RL solutions in the context of improving cryptocurrency operations in Bitcoin and Ethereum: cryptocurrency exchanges \cite{Schnaubelt2022}, trading \cite{Gort2022, asgari2022}, portfolio management \cite{huang2023}, and for example, detecting specific agent behaviors \cite{Schnaubelt2022}, etc. Some new works combine blockchain technologies and RL to improve even economic and social systems, e.g., the healthcare system \cite{Lakhan2023}.

However, these works do not directly address the broadcasting problem in global peer-to-peer networks, namely Bitcoin and Ethereum, and do not pay much attention to the overall environmental impact.

The work most closely related to ours \cite{Yang2019} has already stated that existing blockchain routing schemes based on randomization or fixed-order broadcasting struggle to cope with network dynamics and, in some cases, cannot avoid poorly connected or even malicious nodes when routing changes are made in real-time. In the case of wireless sensor networks, they showed that RL can improve the self-adaptability of the routing scheme by dynamically selecting more reliable and efficient routing channels \cite{Yang2019}. We extended this idea to Bitcoin and Ethereum networks and showed that a well-trained RL agent can efficiently re-prioritize initially randomized broadcasting orders to achieve network consistency earlier and reduce possible traffic redundancy.

%

\section{Methods}
\subsection{The proposed block propagation protocol}
As mentioned earlier, the standard blockchain propagation scheme (BPP) in the Ethereum blockchain implies that a sending node forwards a new block to randomly selected $\sqrt{N}$ neighboring nodes after verifying the block head information. However, this selection may not be optimal in terms of achieving consistency quickly. In some cases, a neighboring node may have a poor direct connection to the sending node, relatively high latency, or even be a black hole node \cite{Yang2019}. In such cases, it may be beneficial to send the block to another neighbor known to have better latency (as well as other important metrics), and the block will eventually reach the neighboring node from a different direction.

Given this, the proposed improvement involves employing a RL agent trained to find the best combination of $\sqrt{N}$ neighboring nodes based on network dynamics and current latency statistics. Thus, the RL agent takes a random list of neighboring nodes and reorders it according to current neighborhood observations. It is important to clarify that the proposed architecture implies a multi-agent approach, where each client of the network has its own agent.

\subsection{Blockchain simulator, network topology and RL-agent}
Our approach is tested on the blockchain simulator developed by Faria et al. \cite{Faria2019}. This simulator is a discrete-event simulator flexible enough to evaluate different blockchain implementations, such as Bitcoin and Ethereum. It follows a stochastic simulation model, capable of representing random phenomena by sampling from a probability distribution. The network model in the simulator is responsible for tracking the state of each node during the simulation, establishing connection channels between nodes, and applying network latency to the messages being exchanged. We extended the network model by implementing BPP broadcasting in a recursive manner and modified network monitoring to collect appropriate statistics. Consequently, the simulator reproduces the standard broadcasting protocol more accurately.

The network latency delay in the simulator is applied depending on the geographic location of the destination and origin nodes. Three geographic locations with corresponding latency distributions were provided out of the box (Ohio, Tokyo, and Ireland), and they showed identical results for average propagation time as in real blockchain, but with a slightly oscillating standard deviation \cite{Faria2019}. Based on the limited number of TCP connections of the average node (approximately 125), we constructed a minimal, but sufficiently large fully connected P2P network architecture of size 150 nodes. Each geographical location consisted of 50 nodes, with half of them marked as miners with equal hash rate (this is a prerequisite for proper simulation).

\begin{figure}
    \centering
    \includegraphics[width=1.0\linewidth]{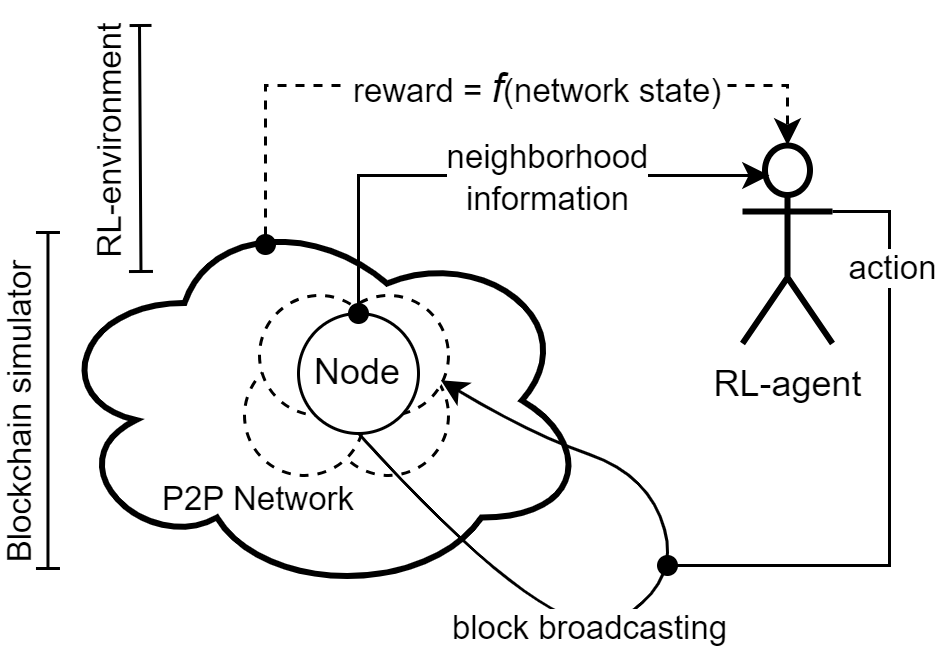}
    \caption{RL-agent and Blockchain simulator integration}
    \label{fig2}
\end{figure}

Next, we constructed an OpenAI Gym RL-environment \cite{Brockman2016} and integrated a simulator instance so that it can be independently executed at each step of training after a RL agent performs its action (Fig. \ref{fig2}). We also modified the node model of the simulator so that it requests the RL agent to reorder the node connections list before starting broadcasting. Thus, the active RL agent might be involved (or not) independently during the simulation.

Based on the tracked information in the simulator, we represent the dynamic network state ($s$) from a node's perspective as an ordered set of neighbor latencies (corresponding to the node connections list) that change after each communication step. For technical simplicity, we use delay estimates obtained from the transaction propagation process, which is performed before the block propagation phase in the simulation. We define an action function as an ordering function over the set of connections between the node and its direct neighbors. According to BPP, each node, by default, randomizes its list of connections and sends full blocks only to $\sqrt{N}$ neighbor nodes, and technically, each action ($a$) represents the corresponding order for the node’s connection list.

To train the RL agent, we employed the Proximal Policy Optimization (PPO) approach. PPO is a family of policy optimization methods that use multiple epochs of stochastic gradient ascent to perform each policy update. These methods offer the stability and reliability of trust-region methods but demonstrate better overall performance and applicability in more general settings\cite{Schulman2017}. There is strong evidence that PPO-based agents are capable of solving pathfinding tasks and are adept at adjusting such heuristics\cite{Skrynnik2022}. Thus, we utilized the OpenAI Gym environment\cite{Brockman2016} along with the Stable-Baselines3 PPO\cite{ppo2021} with mostly default hyperparameters.

\subsection{Key metrics and reward function}
Block propagation time is the most critical metric for blockchains as it determines the fork rate\cite{Gervais2016, Klarman2018} and also limits the frequency of payments. For BPP in Ethereum, the block propagation time is predominantly influenced by the transmission time of the full block\cite{Ma2022}, which can be expressed as:
$$t_{bpp}=l+size\cdot (b^{-1}+t_{proc}),$$
where $t_{bpp}$ indicates block propagation time or total transmission delay; $l$ indicates network latency; $size$ indicates block size; $t_{proc}$ indicates block processing delay; $b$ denotes bandwidth.

Hence, block propagation time should be considered the most important metric when comparing block propagation schemes. Due to the continuous block mining and propagation process, we opt to estimate an average block propagation time over the blocks that were successfully propagated to at least 50\% of nodes in the network (synchronized blocks) and consider this metric as a practical approximation of network consistency – \textbf{synchronization time}.

The other important metric is the probability of blockchain forks or uncle blocks. A fork occurs whenever there are two different valid blocks at the same block height competing to form the longest blockchain. This occurs under normal conditions when two miners solve the proof algorithm within a short period of time from each other. In other words, blockchain forks occur as a result of propagation delays in the global network. A faster block mining time would make transactions faster but lead to more frequent blockchain forks, whereas a slower block time would decrease the number of forks but make settlement slower. Therefore, faster block propagation time reduces the probability that the older block wins the race.

As the probability of forks depends on the block propagation time as well as on the frequency of block introduction (mining in Bitcoin or forging in Ethereum), to practically compare block propagation schemes, we decide to estimate the number of introduced blocks that were successfully propagated to over 50\% of nodes (synchronized) in the network – \textbf{synchronized blocks rate}.

As a proxy for the propagation complexity and the environmental impact, we decide to count the number of messages that are broadcasted through the network. Thus, the third of our metrics is the amount of messages that were emitted to successfully propagate the introduced blocks to over 50\% of nodes in the network – \textbf{messages per synchronized block}.

Based on these metrics, we have designed a reward function for our RL-agent:

$$reward=\frac{\textstyle synchronized{\:}blocks{\:}rate}{\textstyle synchronization{\:}time}$$

\section{Results}

Since we intended to test whether a well-trained RL agent can efficiently re-prioritize broadcasting order and improve the default block propagation protocol by doing so, we ran a series of fixed-duration Ethereum blockchain simulations and evaluated the key metrics mentioned above. To fairly compare BPP with the RL-agent and without it, we performed $k=1000$ simulations with a 60-second duration. This duration is minimally sufficient to introduce up to 5 blocks within a single simulation with comparatively less computation time and overall execution time.

At each of the $k$ iterations, one simulation with and one simulation without the RL agent was run using the same random seed and network topology. We then recorded key metrics: synchronization time, synchronized blocks rate, and messages per synchronized block.

\begin{figure}
    \centering
    \includegraphics[width=1.05\linewidth]{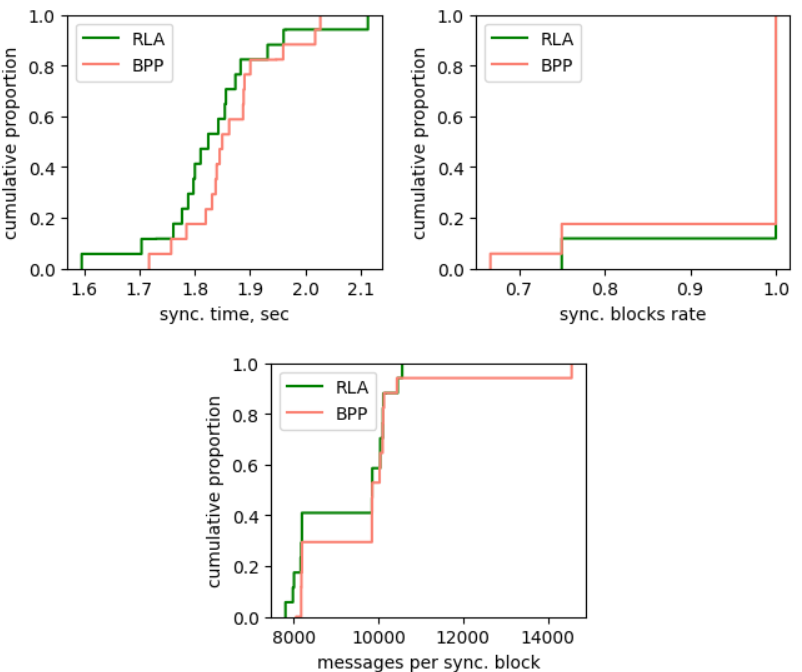}
    \caption{Main simulation results}
    \label{fig3}
    \begin{tablenotes}
      \small
      \item Note. The results of 1000 simulations performed for the default block-hash propagation (BPP) and RL-agent-based block broadcasting (RLA) are depicted in the figures. These figures represent the cumulative fraction of empirical values (ECDF): the higher and farther left the line, the lower the average value. The difference is statistically significant ($p < .001$) using an unpaired Wilcoxon rank sum test with continuity correction.
      
    \end{tablenotes}
\end{figure}

Simulation results confirmed that RL-based broadcasting is able to achieve statistically significant improvements of network dynamics (Fig. \ref{fig3}). On average over 1000 simulations, synchronization time was reduced by 1.7\%, and synchronized blocks rate was increased by 3.4 percentage points. The total number of messages per synchronized block was reduced by 5.0\%.

To estimate the environmental impact of proposed improvements, it is necessary to consider various emission factors relevant to the average Ethereum block and the transmission of related messages via the TCP protocol and Internet infrastructure.

Firstly, there is a direct correlation between the quality of service (QoS) in the TCP communication network and its carbon emissions. In particular, carbon emissions escalate as QoS deteriorates\cite{Habib2013ComparingCP}. This increase is attributed to the growing number of packets generated and their subsequent transmissions via an unstable network. An examination of a modestly sized network, consisting of 100 nodes operating under varied traffic loads and maintaining a minimum of 80\% QoS, reveals an average emission rate of 0.156 gCO2eq. per byte\cite{Usman2017}. This figure is a yearly average based on natural gas as the primary electricity source.

However, deriving a more precise estimate is challenging due to various hardware in use\cite{Usman2017}. Reference data indicates that the peak energy consumption of modern Ethernet switches, which primarily function to route packets, stands at approximately $4.42E{-09}$ gCO2eq. per byte\cite{He_2019, Vishwanath2015}.

Secondly, the size of the average Ethereum block has shown a consistent upward trend. Data from February 2024 indicates that the daily average block size has reached 202,664 bytes, a significant increase from 106,063 bytes a year prior and 25,117 bytes five years earlier\cite{ycharts}. Taking an average block size of 154,363 bytes and assuming that broadcasting one block equates to transmitting one message — without considering additional technical communications and seamless infrastructure nodes of the global Internet — it results in emissions of $6.82E{-04}$ gCO2eq. per message per node. Considering Ethereum's protocol, which introduces a new block every 12-15 seconds to be broadcasted across billions of nodes globally, the potential for emission reduction becomes evident.

By implementing proposed improvements that achieve as much as a 5\% reduction in message transmission within our network model of 150 directly connected nodes, the total carbon emissions could be reduced by approximately $6.82E{-04} \times 485.125 = 0.331$ gCO2eq. during each broadcasting phase. Additionally, the reduction in synchronization time should also lead to a reduction in energy consumption and carbon emissions.

\section{Limitations and future work}
 Using blockchain simulations in experiments is a trade-off between realism and control. Simulations provide a cost-effective, controlled, and safe environment for research and experimentation, but they come with inherent limitations in replicating the complexity of real-world blockchain networks. For instance, simulations can never fully replicate the complexity and dynamics of a real blockchain network; they often rely on simplifications and assumptions, such as uniform network conditions or idealized node behaviors; simulating large-scale blockchain networks with thousands or millions of nodes can be computationally intensive and may not be feasible due to hardware and software constraints etc.
 
 Since we wanted to make not only a controllable and reproducible but also a fair comparison between the default and our improved block propagation protocol, we constructed a simulation-based experimental design. However, in future work we plan to build a small real blockchain testbed, and also consider deeply exploring the proposed approach implementation for other Ethereum-like networks. We also plan to continue experimenting with increasing the degrees of freedom of the solution. For example, by extending the agent's actions with the ability to determine the number of nodes to broadcast the block, instead of being limited by the Ethereum protocol.

\section{Conclusion}
In this paper, we have attempted to address the environmental issues associated with blockchain-based P2P payment networks and explicitly discuss the environmental implications of the proposed improvements. We believe that every small improvement in blockchain protocols and its key metrics should also be considered in terms of energy consumption, data flow, and network load. Therefore, in this work, we:
\begin{enumerate}
  \item Developed an improved Ethereum blockchain peer-to-peer network propagation scheme that involves a RL agent to efficiently re-prioritize the broadcast order based on real-time transport layer network information.
  \item Enhanced an event-driven blockchain simulator by involving a RL agent and created a simulator-based integrated RL learning environment.
  \item Validated the proposed scheme improvement on a number of simulations and showed that the approach reduces the average block propagation time while maintaining a good blockchain synchronization rate.
  \item Emphasized the environmental impact and estimated the required number of broadcast messages before the blockchain network achieved its consistency.
\end{enumerate}

\section*{Code and Data availability}
All data and algorithms used can be found in the public repository: https://github.com/ellariel/eth-broadcast-protocol.

\bibliographystyle{IEEEtran}


%

\end{document}